\documentclass[number,sort&compress,twocolumn,3p]{elsarticle}
\usepackage[binary-units=true,separate-uncertainty=true]{siunitx}
\usepackage{lineno,hyperref}
\usepackage{tikz}

\begin{document}
\begin{frontmatter}
  \title{Status of the CMS Phase~I Pixel Detector Upgrade}

  \author{S.~Spannagel}
  \ead{simon.spannagel@desy.de}
  \author{ on behalf of the CMS Collaboration}
  \address{Deutsches Elektronen-Synchrotron, Notkestr. 85, 22607 Hamburg, Germany}

  \begin{abstract}
    A new pixel detector for the CMS experiment is being built, owing to the instantaneous luminosities anticipated for the Phase~I Upgrade of the LHC.
    The new CMS pixel detector provides four-hit tracking while featuring a significantly reduced material budget as well as new cooling and powering schemes.
    A new front-end readout chip mitigates buffering and bandwidth limitations, and comprises a low-threshold comparator.
    These improvements allow the new pixel detector to sustain and improve the efficiency of the current pixel tracker at the increased requirements imposed by high luminosities and pile-up.
    This contribution gives an overview of the design of the upgraded pixel detector and the status of the upgrade project, and presents test beam performance measurements of the production read-out chip.
  \end{abstract}

  
\end{frontmatter}


\section{Introduction}

The present CMS pixel detector is a hybrid silicon pixel detector and constitutes the innermost component of the CMS tracking system~\cite{CMS-TDR-tracker,CMS-TDR-tracker-add}.
With 66 million readout channels distributed over three barrel layers and two disk layers on either side, it is the experiment's crucial component for the primary and secondary vertex location, b-tagging, and event selection in the high-level trigger.
It has been designed for an instantaneous luminosity of $\mathcal{L} = \SI{1e34}{\centi\meter^{-2}\second^{-1}}$ and performs well under these conditions with single hit efficiencies above \SI{99.5}{\percent}, and a primary vertex resolution of around \SI{20}{\um} for minimum bias events with more than 50 tracks~\cite{tracker-reco}.

However, based on the strong performance of the LHC accelerator, it is anticipated that peak luminosities of two times the design luminosity are likely to be reached before 2018 and probably significantly exceeded in the so-called Phase~I period until 2022.
At this higher luminosity and increased hit occupancies the current CMS pixel detector would be subject to severe dead time and inefficiencies introduced by limited buffers in the analog read-out chip and effects of increased radiation damage in the sensors.

Therefore a new pixel detector is being built and will replace the current detector in the extended year-end technical stop in 2016.

\section{The Phase~I Pixel Detector}

The main objective of the new CMS pixel detector is the maintenance and improvement of the physics performance at higher instantaneous luminosities and around 50 simultaneous inelastic collisions per bunch crossing (pile-up)~\cite{pixel-tdr}.
The increased occupancy at these conditions necessitates new front-end electronics.

Constraints for the new detector are the existing mechanical envelope of the present system as well as the services from the detector patch panel outwards.
In order to operate the new pixel detector, a new powering scheme and faster data links are required.

The tracking performance is improved by providing four barrel and three disk layers for tracking, and by significantly reducing the overall material budget of the detector.
The exchange is foreseen for the extended year-end shutdown of the LHC 2016/2017.

\paragraph{Detector Geometry and Services}

\begin{figure*}[btp]
  \center
  \resizebox{\textwidth}{!}{\tikzset{%
  >=latex
}
\begin{tikzpicture}
  \draw [thick, ->] (0,0) -- (0,4.5);
  \node at (0,5) {$r$};
  \draw [thick, ->] (0,0) -- (13.5,0);
  \node at (14,0) {$z$};
  \node at (-0.25,-0.25) {IP};

  \draw [ultra thick, teal] (0,0.75) -- (6.85,0.75) coordinate (l1end);
  \node [align=right] at(-0.75,0.75) {\SI{30}{\milli\meter}};
  \draw [ultra thick, teal] (0,1.7) -- (6.85,1.7) coordinate (l2end);
  \node [align=right] at(-0.75,1.7) {\SI{68}{\milli\meter}};
  \draw [ultra thick, teal] (0,2.725) -- (6.85,2.725) coordinate (l3end);
  \node [align=right] at(-0.875,2.725) {\SI{109}{\milli\meter}};
  \draw [ultra thick, teal] (0,4) -- (6.85,4) coordinate (l4end);
  \node [align=right] at(-0.875,4) {\SI{160}{\milli\meter}};

  \draw [thick, orange, dashed] (0,1.1) -- (6.85,1.1);
  \draw [thick, orange, dashed] (0,1.825) -- (6.85,1.825);
  \draw [thick, orange, dashed] (0,2.55) -- (6.85,2.55);

  \draw [ultra thick, teal] (7.275,1.125) -- (7.275,4.025);
  \node [align=right] at(7.275,-0.25) {\SI{291}{\milli\meter}};
  \draw [ultra thick, teal] (9.9,1.125) -- (9.9,4.025);
  \node [align=right] at(9.9,-0.25) {\SI{396}{\milli\meter}};
  \draw [ultra thick, teal] (12.9,1.125) -- (12.9,4.025);
  \node [align=right] at(12.9,-0.25) {\SI{516}{\milli\meter}};

  \draw [thick, orange, dashed] (8.624,1.5) -- (8.624,3.75);
  \draw [thick, orange, dashed] (11.625,1.5) -- (11.625,3.75);

  \draw [->] (0,0) -- (9.5818,4.5);
  \draw [->] (0,0) -- (13.5,3.7222);
  \draw [->] (0,0) -- (13.5,2.2313);
  \node at (10.5,4.5) {$\eta = 1.5$};
  \node at (14.5,3.7222){$\eta = 2.0$};
  \node at (14.5,2.3) {$\eta = 2.5$};
  
\end{tikzpicture}}
  \caption{Geometry of the Phase~I barrel and forward pixel detectors. Shown is one quadrant of the detector, the layer radii and disk positions are given relative to the interaction point (IP). Dashed lines indicate the layer positions of the present pixel detector.}
  \label{fig:phase1-geometry}
\end{figure*}
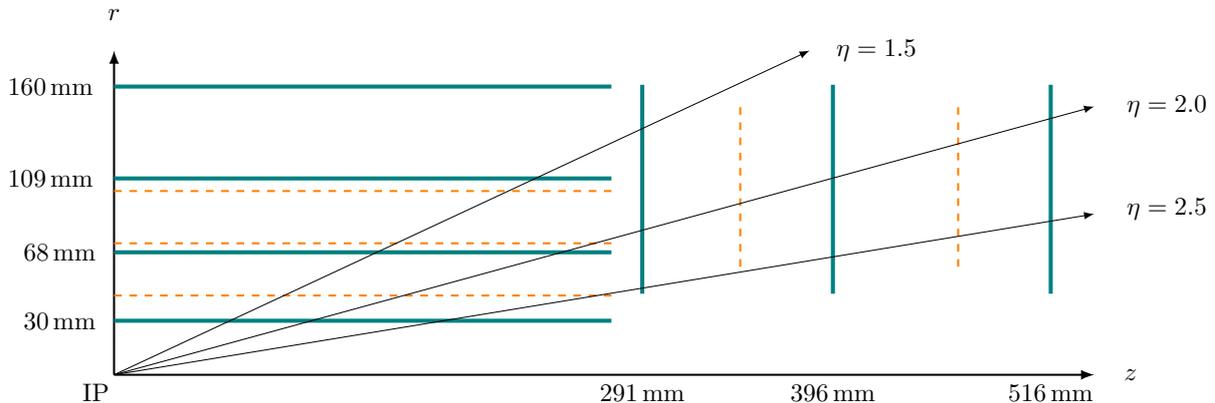

A new beam pipe with an outer diameter of \SI{45}{\milli\meter} has been installed during the first long shutdown of the LHC.
This facilitates the relocation of the innermost pixel layer closer to the interaction point, yielding an improved vertex resolution.
The fourth barrel layer bridges the present gap between the outermost pixel layer and the first layer of the strip tracker, and thus reduces the track extrapolation uncertainty.
The geometries of the present detector and the Phase~I pixel detector are compared in Figure~\ref{fig:phase1-geometry}.
The $z$ axis points along the LHC proton beam, $r$ denotes the radial distance of the barrel layers from the $z$ axis and $\theta$ is the polar angle.
The pseudorapidity is usually favored over the polar angle, and is defined as $\eta = -\ln \tan \theta/2$.

With four barrel layers and $2\times3$ disks, 4-hit coverage up to pseudorapidities of $|\eta| < 2.5$ and a more robust 3-of-4 hit seeding for track candidates is possible.
The total number of channels is almost doubled by the additional detector layers.
The barrel detector granularity increases from about 48 million channels to 79 million channels, while the new forward detector features 45 million channels instead of 18 million for the present detector.

\begin{figure}[h!]
  \centering
  \includegraphics[width=\columnwidth]{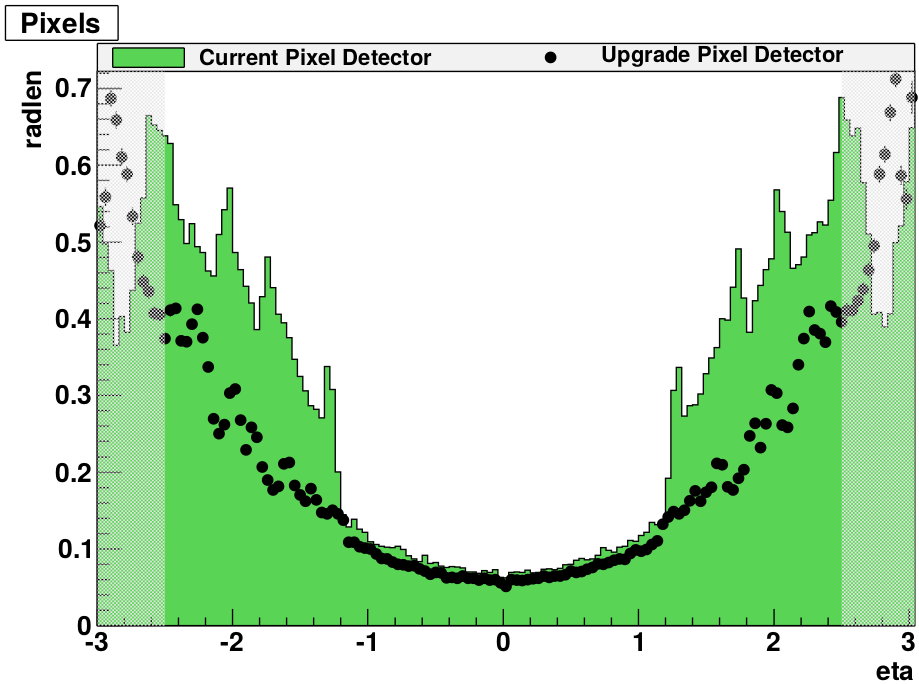}
  \caption{The material budget of the current (green) and the Phase~I (black) pixel detector in units of radiation length as a function of the pseudorapidity $\eta$%
    ~\cite{pixel-tdr}.}
  \label{fig:materialbudget}
\end{figure}

Several measures have been introduced to reduce the amount of material present in the tracking volume as indicated in Figure~\ref{fig:materialbudget}.
A reduced mass produces less photon conversions and multiple scattering from charged particles, and thus yields a better vertex resolution.
A new lightweight support structure made of carbon fiber and graphite compounds reduces the material budget of the mechanical structure, while the 2-phase $\mathrm{CO}_2$ cooling system allows to operate the detector at $T = \SI{-20}{\celsius}$ and further reduces the material due to the low coolant mass and the smaller cooling pipes with $d = \SI{1.6}{\mm}$.
The most significant reduction in material is achieved by relocating many of the service electronics further down on the supply structures, out of the tracking volume.
Together, these changes allow to reduce the total mass despite the additional detector layers.

Owing to constraints on the service interface, existing power supply cables have to be re-used for the operation of a significantly larger detector with its accompanying increased power consumption.
Radiation-hard DC-DC converters have been developed~\cite{dc-dc}, which allow to provide higher voltages over the service cables, and produce the analog and digital supply voltages for the sensors and front-end electronics close to the detector.

\paragraph{Sensors, Front-End Electronics, and Readout}

\begin{figure}[btp]
  \centering
  \includegraphics[width=.95\columnwidth]{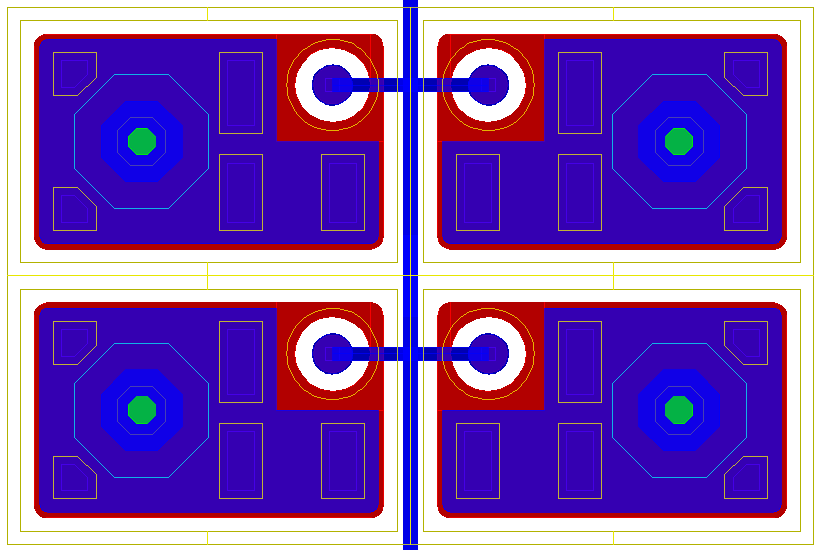}
  \caption{Array of four pixel cells of the barrel pixel sensors. The pixel implants as well as the biasing structure in the center of the array are indicated~\cite{tilman-private}.}
  \label{fig:sensors}
\end{figure}

The sensor design remains unchanged with respect to the present pixel detector.
The \SI{285}{\um} thick \emph{$n^+$-in-n} silicon sensors with either \emph{p}-spray (barrel) or \emph{p}-stop isolation (forward) feature a pixel pitch of \SI{150x100}{\micro\meter} in $r\phi$ and $\eta$, respectively.
In addition to the pixel implants, a biasing grid allows for quality tests of the sensors before being connected to the readout chip (ROC), and is shown in Figure~\ref{fig:sensors}.

The connection to the ROC is established via bump bonds with a diameter of about \SI{25}{\um}.
The ROC is fabricated in a radiation-hard \SI{250}{\nm} analog CMOS technology and presents an advancement of present front-end chip design.
Increased buffers have been integrated in order to mitigate data loss at high occupancies, and a global readout buffer stage allows to reduce the detector dead time.
A faster pixel-cell comparator facilitates in-time thresholds as low as \SI{1.5}{ke}, and the \SI{8}{\bit} on-chip ADC enables digital data transmission at \SI{160}{\mega \bit \per \second}, doubling the effective bandwidth.
A dedicated ROC for the innermost layer is still under design and will comprise a faster pixel readout to further reduce dead time.

The module constitutes the smallest, electrically independent unit of the pixel detector and comprises one sensor and 16 ROCs as well as the high-density interconnect (HDI) and the token bit manager (TBM).
The HDI is a low-material flex print interconnect which is glued onto the sensor backplane, and houses the TBM as well as the routing for data, powering, and trigger signals.
The TBM is responsible for the trigger and token control, and coordinates the readout from the ROCs.
All ROC data are collected by the TBM and transmitted to the receivers outside the experiment at a bandwidth of \SI{400}{\mega\bit \per\second}.
Silicon-nitrite base-strips provide cooling contact and allow fixture of the modules to the mechanical support structure.

\section{Test Beam Qualification of the ROC}

Comprehensive test beam studies for the ROC have been performed at the DESY test beam facility.
The DESY-II synchrotron accelerates positrons up to an energy of \SI{6.3}{\GeV} at a frequency of \SI{1.024}{\MHz}.
This primary beam is converted via bremsstrahlung and pair production into the final test beams with \SI{5}{\percent} momentum spread at rates of a few \si{\kHz}.
Reference tracks are provided by the \textsc{DATURA} beam telescope featuring six planes with MIMOSA26 sensors~\cite{mimosa26} with an intrinsic resolution of \SI{3.4}{\um} and an integration time of \SI{120}{\us}.
Track fitting is performed with the General Broken Lines algorithm~\cite{Blobel2006,Kleinwort2012107}, and track uncertainties at the device under test of about \SI{2.5}{\um} can be achieved even at the comparatively low beam energies provided by the DESY-II synchrotron.

\paragraph{Charge Collection and Tracking Efficiency}

\begin{figure}[btp]
  \centering
  \includegraphics[width=\columnwidth]{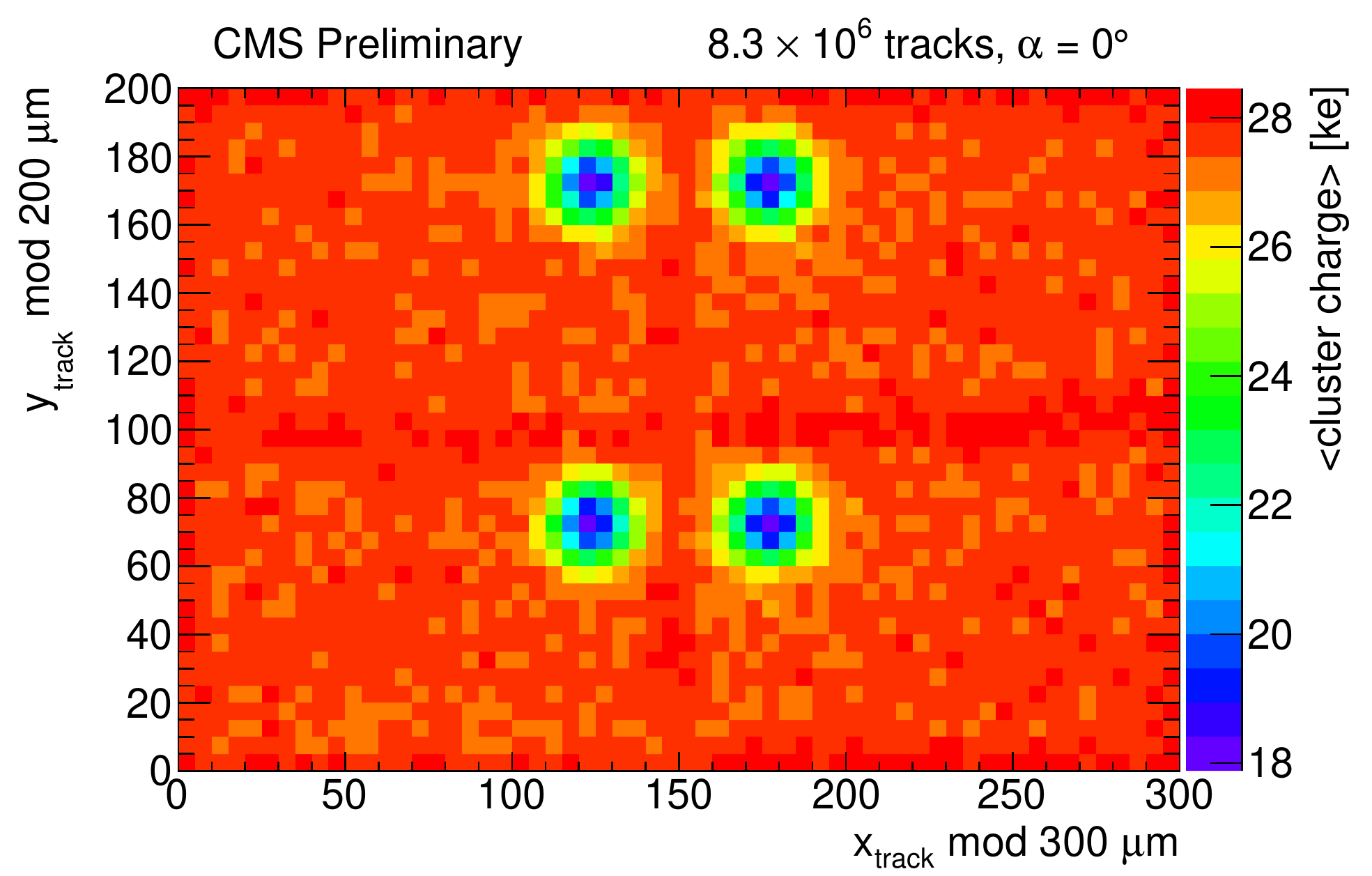}
  \caption{Cluster charge mean as a function of the track impact position within a $2\times2$ pixel array. The charge loss at the biasing dots is clearly visible and amounts to about~\SI{30}{\percent}.}
  \label{fig:cce}
\end{figure}

The charge collection behavior and tracking efficiency of the ROC have been measured as a function of the track impact point within single pixel cells.
Figure~\ref{fig:cce} shows the collected charge as a function of the track impact position for vertical track incidence.
Shown is an array of $2\times 2$ pixel cells with the $x$ and $y$ coordinates denoting the row direction (\SI{150}{\um} pitch) and the column direction (\SI{100}{\um} pitch) of the ROC, respectively.
In order to increase statistics, all tracks from the ROC have been folded into the four pixel cells (Modulo \SI{300}{\um} and \SI{200}{\um}).

The bias dot is clearly visible due to the altered electric field beneath and the thus changed charge collection behavior.
The structure accounts for a charge loss of about \SI{30}{\percent}.

\begin{figure}[btp]
  \centering
  \includegraphics[width=\columnwidth]{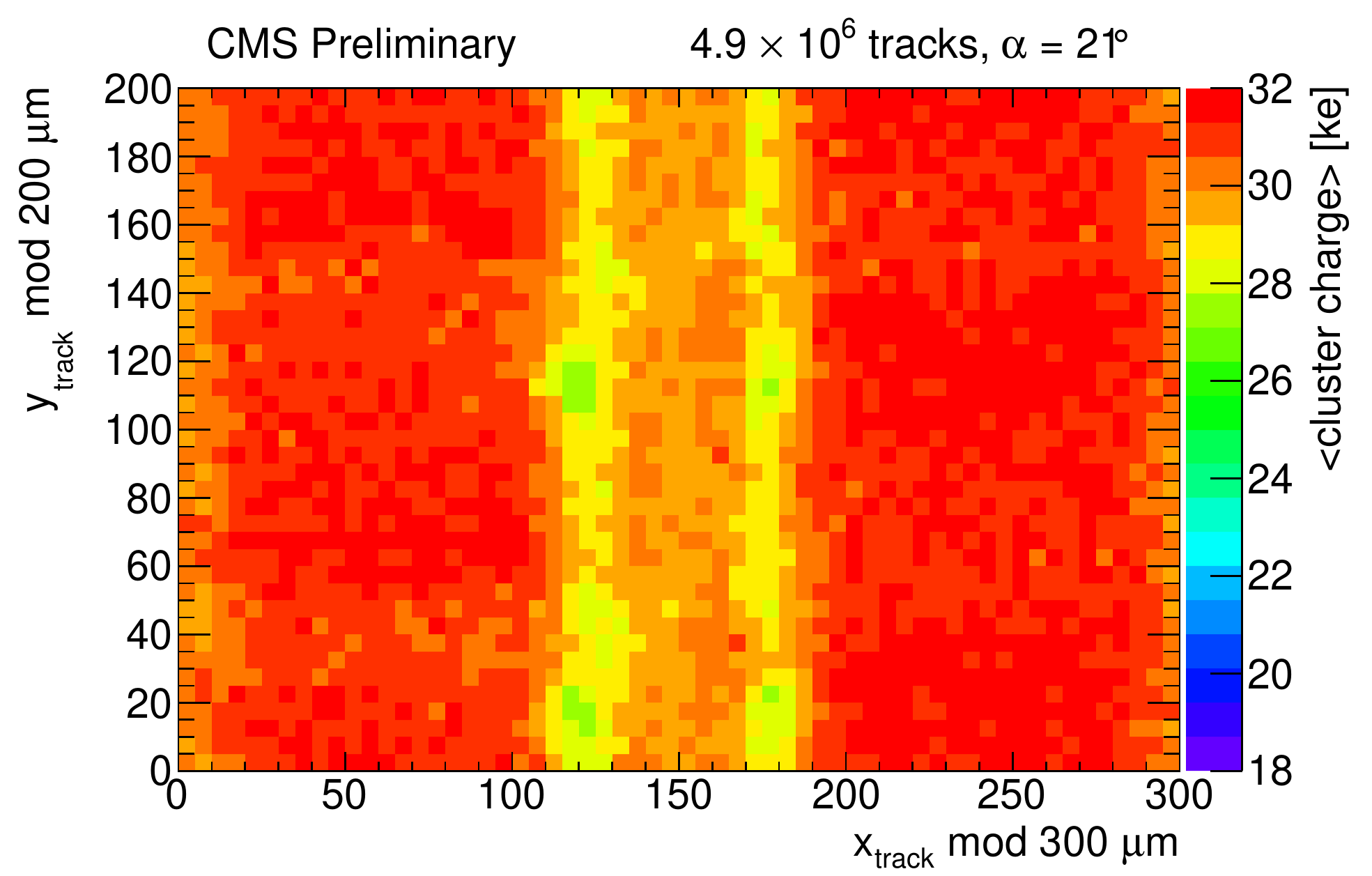}
  \caption{Cluster charge mean as a function of the track impact position, with an incidence angle of \SI{21}{\degree}. The effect of the biasing structure on the charge collection is smeared out and reduced in magnitude to about \SI{10}{\percent}.}
  \label{fig:cce:tilt}
\end{figure}

However, owing to the \SI{3.8}{\tesla} magnetic field in the CMS tracker volume, charges drift through the silicon sensor at an angle of about \SI{20}{\degree}.
This so-called Lorentz angle depends on the applied bias voltage as well as on the radiation damage, and ensures charge sharing between pixel cells.
Thus, a more realistic situation is depicted in Figure~\ref{fig:cce:tilt}, where the track incidence angle is chosen to be \SI{21}{\degree}.
The charge deficiency caused by the biasing grid structure is now smeared out and reduced in magnitude to about \SI{10}{\percent}.

\begin{figure}[btp]
  \centering
  \includegraphics[width=\columnwidth]{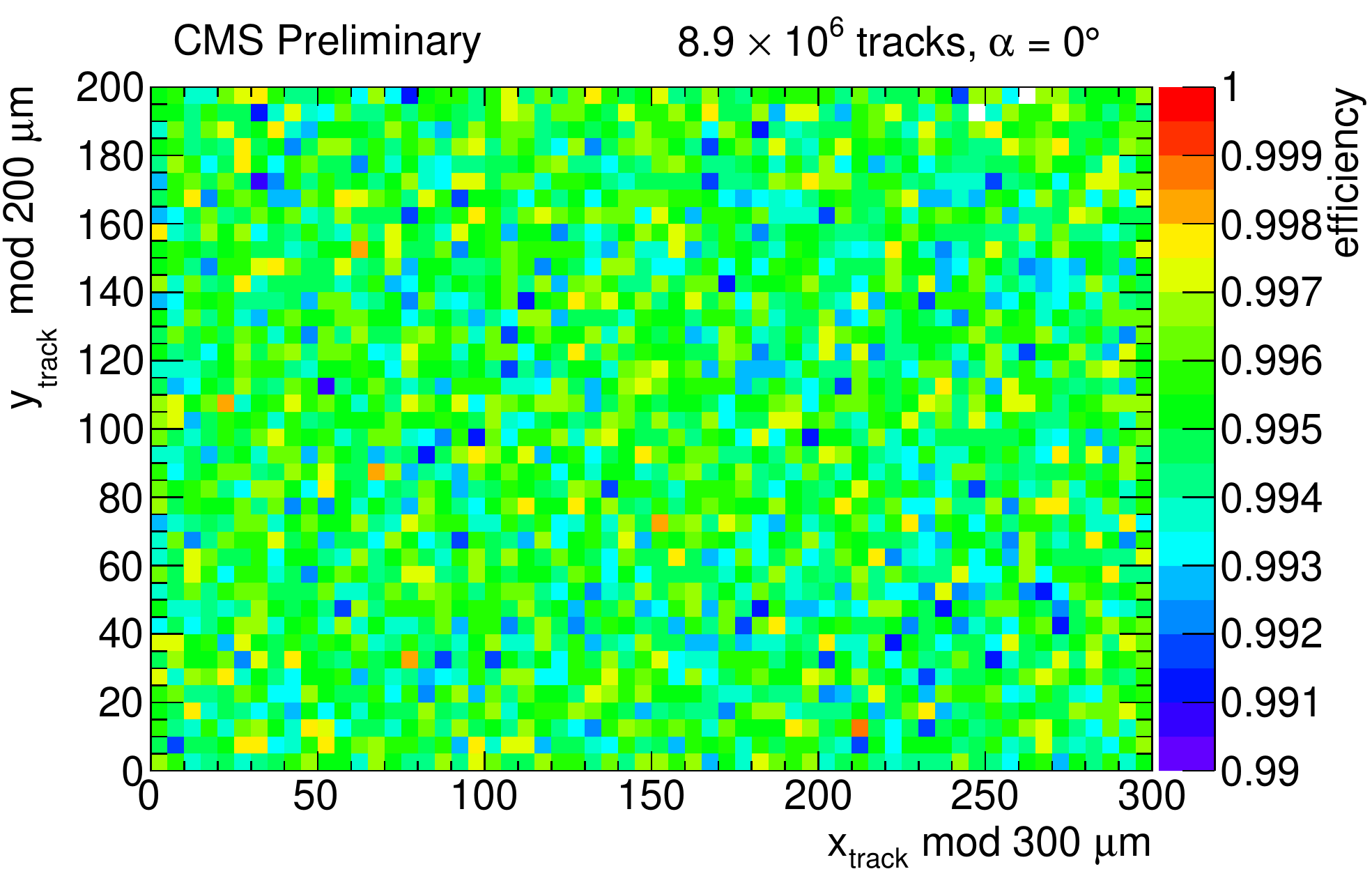}
  \caption{Tracking efficiency as a function of the track impact position at vertical incidence. Shown is an array of $2\times2$ pixel cells. No pattern of particular inefficiencies can be observed.}
  \label{fig:efficiency}
\end{figure}

The tracking efficiency as a function of the track impact point at vertical incidence is shown in Figure~\ref{fig:efficiency}, again representing four pixel cells.
The tracking efficiency is defined as the fraction of reference tracks from the beam telescope with a matched cluster on the ROC over the total number of reference tracks.
There is no structure visible, and the efficiency is well above \SI{99}{\percent} throughout the pixel cell.
The biasing structure has thus no influence on the efficiency, even at vertical incidence where the charge deficiency is most pronounced.

\paragraph{Spatial Resolution and Analog Performance}

\begin{figure}[h!]
  \centering
  \includegraphics[width=\columnwidth]{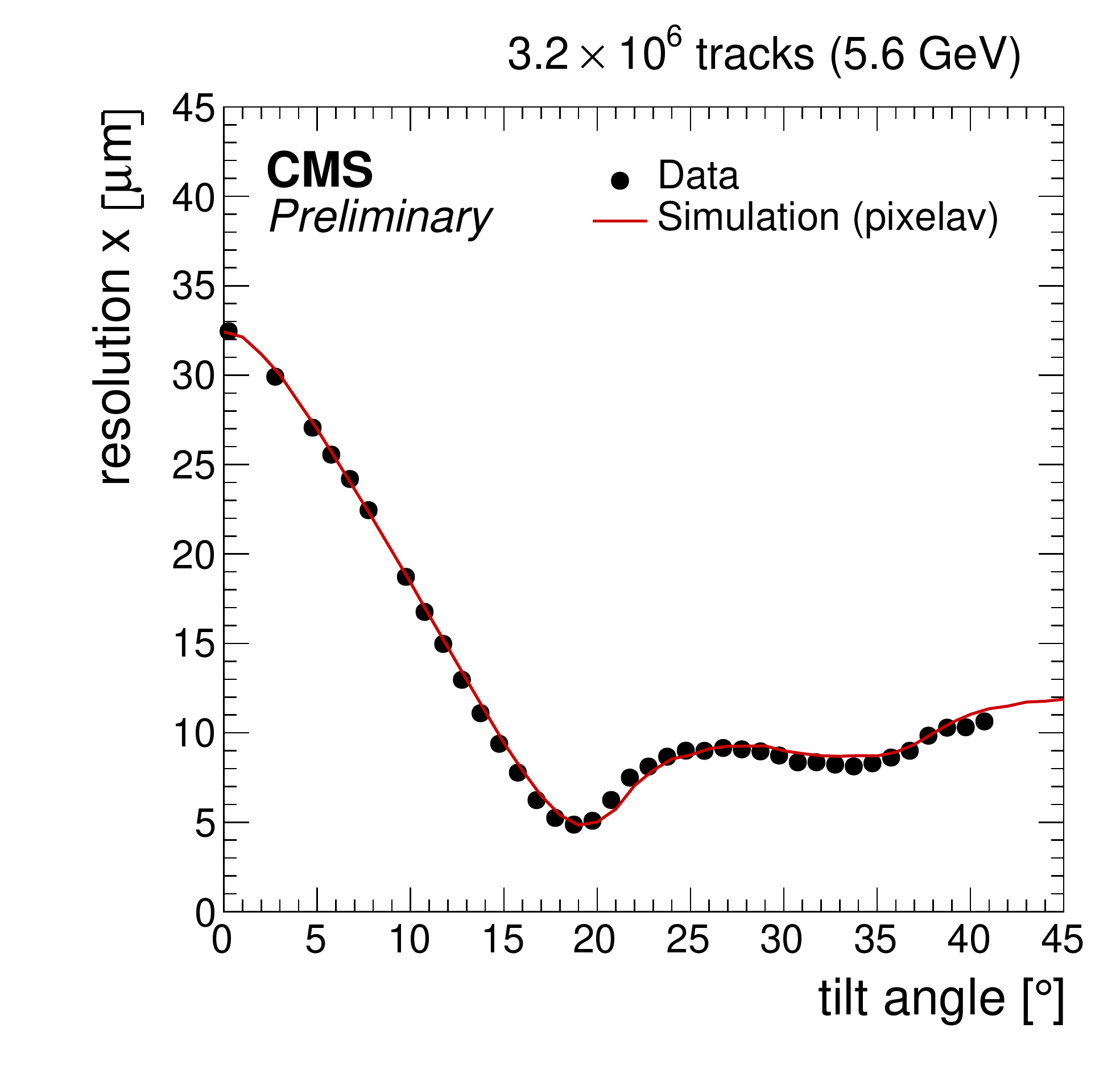}
  \caption{Spatial resolution as a function of the Lorentz angle. Shown is the test beam data along with \textsc{Pixelav} simulations, the pixel pitch is \SI{100}{\um}.}
  \label{fig:resolution}
\end{figure}

The spatial resolution of the pixel detector modules has been measured as a function of the Lorentz angle by rotating the chip assembly in the test beam in order to achieve different track incidence angles as shown in Figure~\ref{fig:resolution}.
The spatial resolution is calculated as the width of the residual distribution of the reference track and the cluster position, with the reference track resolution subtracted quadratically.
The setup is described using the \textsc{Pixelav} simulation software~\cite{Swartz200388,Swartz:687440} and a very good agreement between data and simulation has been achieved.
The best resolution can be measured around the angle of optimal charge sharing between two pixel cells (\SI{19}{\degree} for a pitch of \SI{100}{\um}), and amounts to \SI{4.8}{\um}.

\begin{figure}[btp]
  \centering
  \includegraphics[width=\columnwidth]{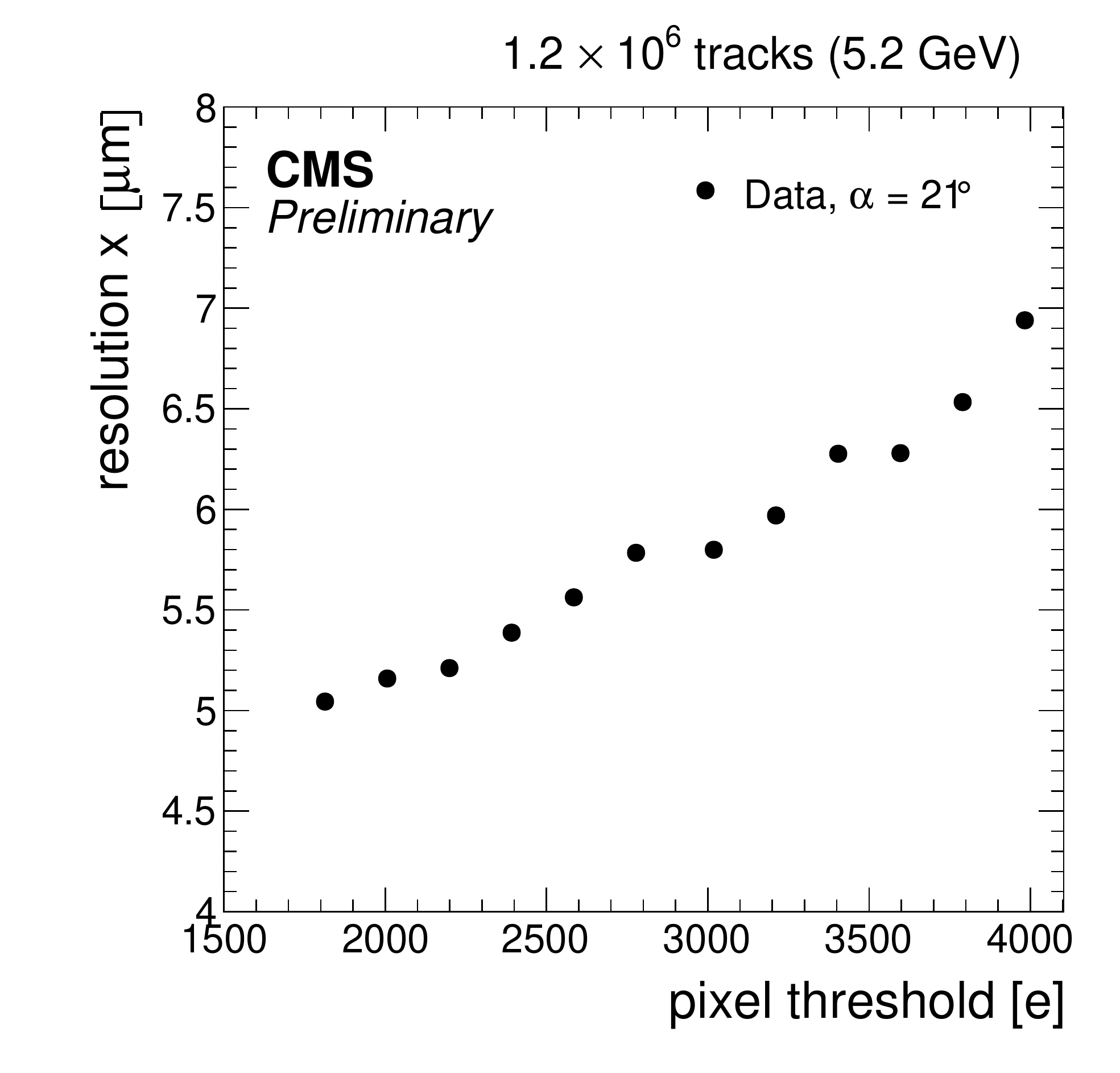}
  \caption{Spatial resolution as a function of the pixel charge threshold. An improvement of about \SI{30}{\percent} can be observed when decreasing the threshold from \SI{4}{ke} to \SI{1.7}{ke}.}
  \label{fig:threshold}
\end{figure}

The improved analog circuitry of the ROC provides a lower absolute and in-time threshold.
While the absolute threshold indicates the total charge at which the comparator triggers the readout of the pixel, the in-time threshold requires the threshold decision to be taken within the same bunch crossing of \SI{25}{\ns}.
The ROC of the current CMS pixel detector allows for an absolute charge threshold of about \SI{2.5}{ke} and an in-time threshold of \SI{3.2}{ke}.
The new ROC is able to record charges as low as \SI{1.5}{ke}.
Owing to a faster comparator and the thereby reduce time walk, the in-time threshold equals the absolute threshold for the new ROC.
The resolution at the track incidence angle of \SI{21}{\degree} has been measured as a function of the pixel threshold, as demonstrated in Figure~\ref{fig:threshold}.
The spatial resolution can be improved by about \SI{30}{\percent} by reducing the pixel charge threshold from \SI{4}{ke} to \SI{1.7}{ke}.

\section{Status of the Production}

The production of the CMS pixel detector modules is split among five production centers.
While the barrel pixel detector is built by institutions in Switzerland, CERN/Taiwan/Finland, Italy, and Germany, the disks are built by a consortium of U.S.\ universities and research centers.
The module production for the detector layers 2--4 started in the second quarter of 2015, and all modules are expected to be delivered to the central integration site at PSI, Switzerland, by the first quarter of 2016.
Due to the special ROC required for layer 1 of the detector, the module production for this part will only start in summer 2016.

A pilot blade system has been installed in CMS during the long shutdown, which allows to operate pre-series production modules of the new Phase~I pixel detector already in realistic conditions in the CMS pixel detector in order to gain experience with the system.

The detector integration will start by the end of 2015, with the testing and commissioning being performed throughout 2016.
Finally, the new detector will be installed in the CMS experiment during the extended year-end shutdown of the LHC in winter 2016/2017.

\section{Conclusion}

The present CMS pixel detector performs well at the current operation conditions, but would be subject to severe efficiency losses at instantaneous luminosities of $\mathcal{L} = \SI{2e34}{\centi\meter^{-2}\second^{-1}}$ as expected from the LHC schedule.
Thus, the detector will be replaced by the new Phase~I pixel detector during the extended LHC winter shutdown 2016/2017.
The Phase~I upgrade system comprises four barrel layers and $2\times3$ disks, a reduced material budget, and a new cooling system.
The front-end electronics have been redesigned in order to withstand the higher particle fluxes, and to improve the overall tracking performance.
Test beam measurements have been conducted in order to qualify and characterize the new ROC, and a pilot system inside the CMS experiment allows to gain experience in the final deployment environment.
Most of the individual components are fabricated, and the module production is ongoing.
During 2016, the modules will be integrated and the full detector commissioned.

\section*{References}
\bibliography{bibliography}

\end{document}